\documentclass[ reprint, aps,]{revtex4-2}
\usepackage{amssymb}
\usepackage{graphicx}
\usepackage{dcolumn}
\usepackage{bm}

\begin{document}

\title{Hubble tension and Reheating: Hybrid Inflation Implications}
\author{K. El Bourakadi}
\email{k.elbourakadi@yahoo.com}
\date{\today }

\begin{abstract}
We investigate a new possible solution to the Hubble constant tension$.$ we
propose a simple resolution to the problem assuming that a first-order phase
transition related to $H_{0}$\ transition occurred in the early Universe.
The early evolution of the Universe is a result of hybrid inflation that has
lasted for a specific period until symmetry breaking takes place. Fitting
our model to measurements from $Planck$ and $SH0ES$ data provides a key
explanation of discrepancies of $H_{0}$ measurements. The quantum
fluctuations calculated in this model have significant results on the
reheating parameters $N_{re}$ and $T_{re}.$ Therefore, new constraints must
be taken into consideration to fit these parameters to recent results.
\end{abstract}

\preprint{APS/123-QED}


\affiliation{ Physics and Quantum Technology Team, LPMC, Ben M’sik Faculty
of Sciences, Casablanca Hassan II University, Casablanca, Morocco } 
\altaffiliation[Also at ]{LPHE-MS Laboratory Department of Physics,
Faculty of Science, Mohammed V University in Rabat, Rabat, Morocco} 

\affiliation{ LPHE-MS Laboratory Department of Physics, Faculty of Science, Mohammed V University in Rabat, Rabat,
Morocco }%
\altaffiliation[Also at ]{LPHE-MS Laboratory Department of Physics, Faculty of Science, Mohammed V University in Rabat, Rabat,
Morocco}






\maketitle
\section{\label{sec:1} Introduction}

The Hubble parameter $H_{0}$ calculated using CMB radiation observations,
puts the $\Lambda CDM$ model in a crisis since it does not agree with the
results obtained by directly measuring the expansion rate today using
supernovae redshift measurements. The Planck satellite which offers the most
accurate observations of temperature fluctuations, polarization, and lensing
in CMB radiation, considers that the $\Lambda CDM$ model predicts the
current value of the expansion rate today to be $H_{0}=67.36\pm \ 0.54\ km\
s^{-1}\ Mpc^{-1}$ \cite{G1}. On other hand, the expansion rate measured from
Cepheids-calibrated supernovae by the $SH0ES$ team \cite{G2}, was found to
be $H_{0}=74.03\pm 1.42\ km\ s^{-1}\ Mpc^{-1}$, several other Hubble rate
measurements including $H0LiCoW$ \ \cite{G3} also show substantial
discrepancies. A popular conclusion among astronomers is that additional
physics outside of the $\Lambda CDM$ model is needed to resolve the tension 
\cite{G4,G4-1}, but no consensus has yet been reached on what those effects
may be. While it is important to continue to look for possible systematic
impacts, the possibility of an extra component of dark energy in the early
Universe could be the reason for the change in the measurement of $H_{0}$.
According to \cite{G7,G7-1} an early scenario could occur thanks to an extra
dynamical scalar field, which behaves like dark energy until a crucial
moment, at which point its energy density quickly decays.

Reheating is the phase at which the transition from inflation to the
radiation era occurs, during this step the production of relativistic
particles take place thanks to the inflaton field decay, reheating could
either involve a perturbative decay of the inflaton field at the end of
inflation \cite{Ki6}, or include a scenario of parametric resonance we call
preheating \cite{Ki7}. The physics of reheating is usually studied
independently on dark energy effects. However, according to \cite{G13}\
inflationary\ quantum fluctuations could be the source behind the expansion
of the Univers described through dark energy, following this results it is
possible to constrain reheating parameters $N_{re}\ $and $T_{re}$ according
to recent data, considering that hybrid inflation behaves like dark energy
until the symmetry breaking occurs.

In the present paper, we consider hybrid inflation as a solution to the
discrepancies of the expansion rate measured from deferents sources. Hybrid
inflation models are described usually by at least two scalar fields, we
will study a model with $\phi $ and $\sigma $ fields, which looks like a
hybrid of chaotic inflation with $V(\phi )=m^{2}\phi ^{2}$ and a
non-inflationary potential with spontaneous symmetry breaking given by $%
V(\sigma )=\left( 1/4\lambda \right) (M^{2}-\lambda \sigma ^{2})^{2}$ \cite%
{G8}. In this model, inflation ends by a first-order phase transition, in
this direction we will investigate the possibility that the phase transition
that occurred in the early universe before recombination, which could be
responsible for the transition of $H_{0}$ from its initial value to its
current value. Finally, we extract information about reheating in terms of
the e-folds numbers $N_{re}$\ and the reheating temperature $T_{re}$ and put
new constraints on this phase studying the effects of the hybrid inflation
parameters on reheating parameters.

Our paper is organized as follows, in the next section, we first recall the
basic notions of the inflationary quantum fluctuations and their effects on
initial measurements of $H_{0}$. In sect. \ref{sec:3}, we discuss the
mechanism of the symmetry breaking that could explain the expansion rate
transition. In sect. \ref{sec:4}, we investigate the constraints on hybrid
inflation parameters from the early evolution of the universe. In sect. \ref%
{sec:5}, we study the reheating phase, and we focus on the duration $N_{re}$
and temperature $T_{re}$ constraints according to recent observations.

\section{\label{sec:2}Initial measurement from quantum fluctuations}

According to the hybrid inflation model proposed in \cite{G9}, which is
given as 
\begin{equation}
V(\phi ,\sigma )=\frac{1}{4\lambda }\left( M^{2}-\lambda \sigma ^{2}\right)
^{2}+\frac{1}{2}m^{2}\phi ^{2}+\frac{1}{2}g^{2}\phi ^{2}\sigma ^{2},
\label{1}
\end{equation}%
during the slow-roll of the inflaton field, $\phi $ moves towards the
critical value $\phi _{c}=M/g$, and the field fluctuate at the minimum of
the potential $\sigma =0$, that takes the form$\ V(\phi )=M^{4}/4\lambda
+m^{2}\phi ^{2}/2$. For $\phi >\phi _{c}$, $\sigma $ has positive effective
mass squared. Inflation occurs while $\phi $ decreases slowly due to the
increasing chaotic term $V(\phi )=m^{2}\phi ^{2}/2$ in Eq. (\ref{1}).

Inflation ends at $\phi =\phi _{c},$ the energy density became dominated by
the false vacuum contribution. When $\phi <\phi _{c}$, $\sigma $ acquires a
tachyonic effective mass and the fields roll towards the true minimum at $%
\phi =0$.

Considering that $\phi >\phi _{c}=M/g,$ a model was considered in \cite{G17}
where they described a rapid rolling of $\sigma $ field\ to realize the
waterfall mechanism which makes inflation ends in a different way, taking
into account that the mass $m$ of the scalar field $\phi $\ is in the order
of the Hubble scale $m\sim H_{0}\sim 1.4\times 10^{-33}eV$. The massless
scalar field will acquire quantum fluctuations during inflation since the
parameter $H_{I}$ satisfies the condition $H_{I}\gg H_{0}$. We decompose the
classical field $\phi $ into homogeneous mode and linear perturbations as

\begin{equation}
\phi (\mathbf{x},t)=\varphi (t)+\delta \phi (\mathbf{x},t)
\end{equation}%
we then make Fourier transform of the linear perturbations $\delta \phi (%
\mathbf{x},t)$ as \cite{G11,G12}

\begin{equation}
\left\vert \delta \phi _{\mathbf{k}}\right\vert ^{2}\simeq \frac{H_{I}^{2}}{%
2k^{3}}\left( \frac{k}{aH_{I}}\right) ^{2m^{2}/\left( 3H_{I}^{2}\right) },
\end{equation}%
knowing that inflation start at $a_{i},$ variations in the real space field
caused by super-Hubble fluctuations for $N=\ln \left( a_{e}/a_{i}\right) $
inflation $e$-folds are expressed as \cite{G13}:%
\begin{eqnarray}
\left\langle \delta \phi ^{2}\right\rangle &\simeq
&\int_{a_{k}H_{I}}^{aH_{I}}\frac{d^{3}\mathbf{k}}{\left( 2\pi \right) ^{3}}%
\left\vert \delta \phi _{\mathbf{k}}\right\vert ^{2}  \nonumber \\
&=&\frac{3H_{I}^{4}}{8\pi ^{2}m^{2}}\left[ 1-\exp \left( -\frac{2m^{2}}{%
3H_{I}^{2}}N\right) \right] ,  \label{eq2}
\end{eqnarray}%
the exponential term which represents the evolution of the homogeneous mode
must be suppressed compared to the field fluctuations \cite{G13}, for that
reason inflation lasts long enough for the exponential term to cancel, which
will lead to $\left\langle \delta \phi ^{2}\right\rangle =3H_{I}^{4}/8\pi
^{2}m^{2}$. Before phase transition, the energy density in the first phase
supported by quantum fluctuations of the inflationary potential $V\left(
\phi \right) $ is roughly

\begin{equation}
V\left( \phi \right) \simeq \frac{1}{2}m^{2}\left\langle \delta \phi
^{2}\right\rangle \simeq \frac{3H_{I}^{4}}{16\pi ^{2}},  \label{Vi}
\end{equation}%
from Eq. \ref{Vi} and the density parameter, one should express the
parameter $H_{I}$\ as

\begin{equation}
H_{I}\simeq \left( \Omega _{\Lambda }\right) ^{1/4}\sqrt{4\pi H_{0}M_{p}},
\label{Hi}
\end{equation}%
here, $M_{p}$ is the Planck mass, and $\Omega _{\Lambda }$\ is the density
parameter associated with a cosmological constant \cite{G1}. Using the
current data from $Planck$, one gets the results in table (\ref{table:1}).

\begin{table*}[tbp]
\caption{The Hubble constant associated with inflation considering parameter
68\% intervals for the base-$\Lambda CDM$ model from Planck $CMB$, in
combination with $CMB$ $lensing$ and $BAO$.}
\label{table:1}%
\begin{ruledtabular}
\begin{tabular}{cccc}
& \footnotesize TT,TE,EE+lowE & \footnotesize TT,TE,EE+lowE+lensing & \footnotesize TT,TE,EE+lowE+lensing+BAO \\ \hline
$H_{0}\left[ km~s^{-1}~Mpc^{-1}\right]$ & $67.27\pm 0.60$ & $67.36\pm 0.54$ & $67.66\pm 0.42$ \\
$H_{I}~\left[ eV\right] $ & $6.017\times 10^{-3}$ & $6.021\times 10^{-3}$ & $%
6.034\times 10^{-3}$ \\
\end{tabular}
\end{ruledtabular}
\end{table*}

\begin{figure}[h]
\includegraphics[width=12cm]{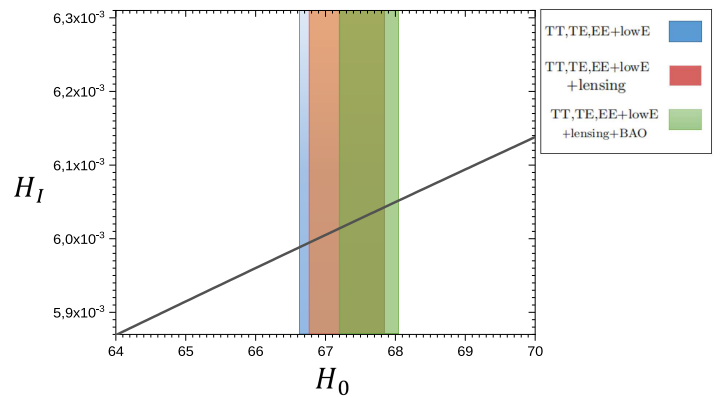}
\caption{Variation of $H_{I}$ as a function of $H_{0}$. The dark energy
density is considered to be $\Omega _{\Lambda }=0.68.$.}
\label{fig:1}
\end{figure}

From the estimations in table \ref{table:1} and the massless inflaton field $%
m\ll H_{I},$ one would consider an extremely large number of $e$-folds as a
requirement from Eq. (\ref{eq2}). As a consequence, the homogeneous mode of
the scalar field is therefore considerably reduced, such that the long wave
fluctuation $\left\langle \delta \phi \right\rangle $\ determines the value
of the classical field.

In Fig. (\ref{fig:1}) the variation of $H_{I}$\ as a function of the Hubble
rate $H_{0}$ is presented, the vertical light blue, pink and green regions
represents Planck's bounds on $H_{0}.$ We observe that $H_{I}$ is an
increasing function that\ shows good compatibility with observation when $%
H_{I}>5.99\times 10^{-3}$ . However, this parameter must obey the condition $%
H_{I}<6.05\times 10^{-5}$ in order to repreduce valid values of $H_{0}$\
according to observations.

\section{\label{sec:3}Secondary measurement from phase transition}

At the moment when the inflaton field $\phi $ becomes smaller than $\phi
_{c}=M/g$, symmetry breaking occurs. then the phase transition caused by the 
$\sigma $ field fluctuations ends inflation. When $m^{2}\phi ^{2}<m^{2}\phi
_{c}^{2}=m^{2}M^{2}/g^{2}$, inflation ends due to the waterfall mechanism.
Thus, any quantum fluctuation will not be produced in this phase since $%
\sigma \rightarrow M/\sqrt{\lambda }$ and $\phi \rightarrow 0$, this will
cause an instantaneous transition and as in the original hybrid inflation
model, inflation will come to an abrupt end \cite{G14,G15}. The potential
energy at the transition is given as \cite{G17}

\begin{equation}
V_{T}\simeq \frac{M^{4}}{4\lambda }=\frac{H_{T}^{4}}{4},  \label{Vt}
\end{equation}%
to explain the expansion rate transition one needs

\begin{equation}
H_{T}\simeq \left( \Omega _{\Lambda }\right) ^{1/4}\sqrt{12~H_{0}M_{p}}.
\end{equation}

The present data from $SH0ES$ team \cite{G2}, provides the results in table (%
\ref{table:2})

\begin{table*}[tbp]
\caption{The Hubble constante associated with the transition considering
differents measurements of the best Estimates of $H_{0}$ Including
Systematics from $SH0ES$ team. }
\label{table:2}%
\begin{ruledtabular}
\begin{tabular}{cccc}
& \footnotesize LMC + NGC 4258 & \footnotesize LMC + MW & \footnotesize %
NGC 4258 + MW  \\ \hline
$H_{0}\left[ km~s^{-1}~Mpc^{-1}\right]$ & $73.40\pm 1.55$ & $73.94\pm 1.58$ & $74.47\pm 1.45$ \\
$H_{T}~\left[ eV\right] $ & $6.142\times 10^{-3}$ & $6.164\times 10^{-3}$ & $%
6.186\times 10^{-3}$ \\
\end{tabular}
\end{ruledtabular}
\end{table*}

\begin{figure}[h]
\includegraphics[width=11.5cm]{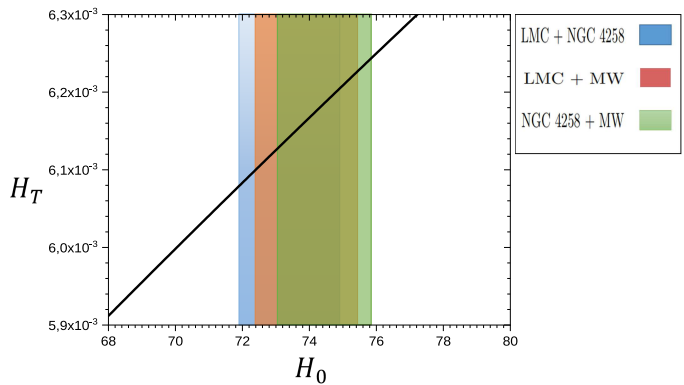}
\caption{Variation of $H_{T}$ as a function of $H_{0}$. The observational
data were taken from $SH0ES$ team.}
\label{fig:2}
\end{figure}

Fig. (\ref{fig:2}) represents the variation of $H_{T}$ according to $H_{0},$
the value of the Hubble rate is restricted between $71.85~km\ s^{-1}\
Mpc^{-1}<H_{0}<75.92~km\ s^{-1}\ Mpc^{-1}$\ from observations. We observe
that the parameter $H_{T}$ should be bounded as $6.08\times
10^{-3}~eV<H_{T}<6.24\times 10^{-3}~eV$ in order to obtain compatible values
from the $SH0ES$ team measurements of $H_{0}.$

When we consider $M>H_{I},$ we can automatically conclude that the parameter 
$\lambda $\ is bounded as

\begin{eqnarray}
\lambda &>&\frac{H_{I}^{4}}{H_{T}^{4}},  \nonumber \\
\lambda &\gtrsim &1\ \ \ \ \ .
\end{eqnarray}

\section{\label{sec:4}Constraints from early evolution}

Since inflation lasted for an extremely long period according to the above
analysis, the early evolution could therefore be a natural consequence of
cosmic inflation. The stage of inflation is defined\ at large $\phi $, where
the effective potential $V(\sigma ,\phi )$ is at the local minimum $\sigma
=0 $, therefore, $V(0,\phi )=M^{4}/4\lambda +m^{2}\phi ^{2}/2$.

When we assume $m\ll H_{I},$ the condition from the lower bound on the
inflationary quantum fluctuation $\left\langle \delta \phi ^{2}\right\rangle
=3H_{I}^{4}~/8\pi ^{2}m^{2}$ can be given as

\begin{equation}
\left\langle \delta \phi ^{2}\right\rangle \gg \frac{3H_{I}^{2}}{8\pi ^{2}}.
\end{equation}

The amplitude of density perturbations produced in the model $V\sim
m^{2}\phi ^{2}/2$ can be estimated as \cite{G17,G1}

\begin{eqnarray}
\mathit{P}_{R}^{1/2} &=&\frac{16\sqrt{6\pi }}{5}\frac{V^{3/2}}{M_{p}^{3}%
\frac{\partial V}{\partial \phi }}\sim \frac{2\sqrt{6\pi }}{5}\frac{m~\phi
^{2}}{M_{p}^{3}}, \\
\mathit{P}_{R} &\simeq &A_{s}=2.196_{-0.06}^{+0.051}\times 10^{-9}
\end{eqnarray}%
in this case, the amplitude depends on the field $\phi >\phi _{c}$, as a
consequence the constant $g$\ and the masses $M,m$ should satisfy the
condition

\begin{equation}
\frac{m}{M_{p}^{3}}\frac{M^{2}}{g^{2}}<2.69\times 10^{-5},
\end{equation}%
the constant $g$ with the bare masses could practically have any value, as
long as they satisfy the constraint above.

\section{\label{sec:5} Reheating Constraints from early evolution}

Reheating is a phase that occurs in the early universe, at which, the
creation of elementary particles and their decay takes place. In fact, it
has been proposed in\ \cite{K10,K11-1} that during the first stage of
reheating (preheating), phenomena like primordial gravitational waves can be
produced and the density spectrum of gravitational waves produced during
this stage can satisfy constraints from the latest $Planck^{\prime }s$ data.
Knowing that preheating is characterized by parametric resonance, the
created particles can only decay and thermalize with a final temperature
called reheating temperature in the last stage of reheating, the decay rate
of the inflaton oscillations parametrized with $\Gamma $ \cite{K9,K12}, is
added as a friction term to the inflaton equation of motion (EoM) $\ddot{\phi%
}+3H\dot{\phi}+V\left( \phi \right) =0$\textbf{, }the decay rate\textbf{\ }%
describing the energy transferred to new particles is given by the formula $%
\Gamma =$ $\Gamma \left( \phi \rightarrow \chi \chi \right) +$ $\Gamma
\left( \phi \rightarrow \psi \psi \right) .$ In our model reheating must
occur considering the condition $\phi \gtrsim \phi _{c},$\ that makes our
model described by a chaotic shape of potential $V\sim m^{2}\phi ^{2}/2.$%
\textbf{\ }We can extract\ information about reheating in terms of the $e$%
-folds numbers $N_{re}$\ and the reheating temperature $T_{re}$ considering
the time observed CMB modes crossed beyond the Hubble radius till the
present time \cite{K5,K11}

\begin{equation}
N_{pre}=\left[ 61.6-\frac{1}{4}\ln \left( \frac{V_{end}}{H_{k}^{4}}\right) -N%
\right]
\end{equation}

\begin{equation}
T_{re}=\left[ \left( \frac{43}{11g}\right) ^{\frac{1}{3}}\frac{a_{0}T_{0}}{k}%
H_{k}e^{-N}\left( \frac{3^{2}\cdot 5V_{end}}{\pi ^{2}g}\right) ^{-\frac{1}{%
3(1+\omega)}}\right] ^{\frac{3(1+\omega)}{3\omega-1}},
\end{equation}%
next, we express the two inflationary parameters $N$ and $V_{end}$\ as a
function of the scalar spectral index \cite{K6}

\begin{eqnarray}
N &=&\frac{2}{1-n_{s}} \\
V_{end} &=&6\pi ^{2}M_{p}^{4}A_{s}\left( \frac{\left( 1-n_{s}\right) }{2}%
\right) ^{2},
\end{eqnarray}%
the parameter $H_{k}$ can be written in terms of the chaotic potential \cite%
{K7}

\begin{equation}
H_{k}=2\pi \sqrt{A_{s}M_{p}^{4}\left( \frac{V^{\prime }}{V}\right) ^{2}},
\end{equation}%
after calculation we make the transition $\phi ^{2}\rightarrow \left\langle
\delta \phi ^{2}\right\rangle ,$ knowing that $\left\langle \delta \phi
^{2}\right\rangle =3H_{\inf }^{4}/8\pi ^{2}m^{2}$

\begin{equation}
H_{k}=\frac{\pi ^{2}m M_{p}^{2}}{H_{I}^{2}}\sqrt{\frac{128}{3}A_{s}}.
\end{equation}

\begin{figure*}[tbp]
\includegraphics[width=18cm]{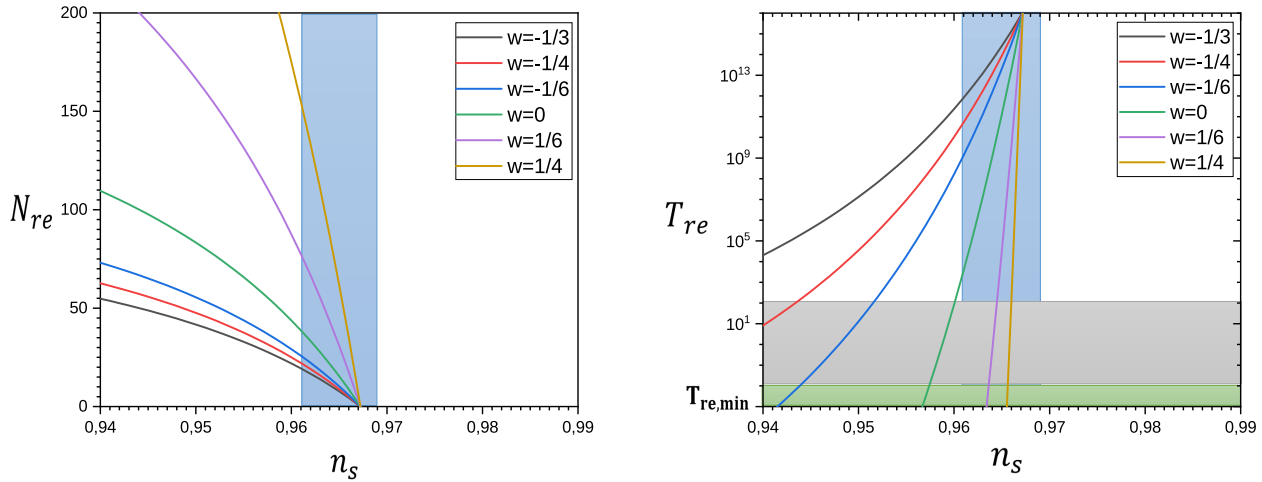}
\caption{Variation of $N_{re}$ and $T_{re}$ as a function of the spectral
index $n_{s}$. The observational data were taken from $Planck^{\prime }s$
results.}
\label{fig:3}
\end{figure*}

Fig. (\ref{fig:3}) presented the variation of reheating parameters $N_{re}$\
and $T_{re}$ as a function of the spectral index $n_{s}$, the equation of
state (EoS) parameter $\omega$ takes different values each one with a
specific color, the EoS varies from $-1/3$ to $1/4,$\ knowing that $%
\omega=1/3$ cannot present any predictions for $N_{re}~$and $T_{re}.$\ The
blue vertical regions show the values of the scalar spectral index released
by $Planck^{\prime }s$\ data, and the horizontal regions for the reheating
temperature plot, represent the temperatures below the electroweak scale $%
\left( T<100GeV\right) $\ and below the big bang nucleosynthesis scale $%
\left( T<10MeV\right) $ respectively. Note that for the reheating duration $%
N_{re}$, we define the instantaneous reheating by the limit $%
N_{re}\rightarrow 0$\ at the point where all the lines converge, our chaotic
potential of inflation shows compatibility with observations for all
diferent values of $\omega,$ in addtion to that the case of $\omega=1/4$\
gives good consistency even for extremely higher $e$-folds number. On the
other hand for the case of the reheating temperature $T_{re}$, An
instantaneous reheating leads to the maximum temperature $T_{re}\sim
10^{16}GeV$. Our model chooses to be compatible for all cases of $\omega$
espacially for $\omega=1/4$ that can reach the lowest possible values of the
reheating temperature $T_{re,\min }>4MeV$ \ \cite{K8}$.$

At the end of hybrid inflation, the behavior of the fields could describe
the explosive preheating with a production of $\phi $ and $\sigma $
particles \cite{G8}. While in \cite{K11} the behavior of particle production
was studied considering an extra field $\chi $ coupled to both $\phi $ and $%
\sigma $ fields. Particle production in $\chi $ field that interacts with
the fields $\phi $ and $\sigma $ require amplified fluctuations which allows
an explosive production of $\chi $-particles, this means we are in the broad
parametric resonance region, the expansion of the universe plays an
important role in ending the parametric resonance regime, particle
production will be ended by the redshifted modes that will fall out of the
resonance band because of the expansion.

\section{\label{sec:6}Conclusion}

In this work we considered hybrid inflation as a solution to the Hubble
tension, we have discussed the possibility that the transition of Hubble
rate parameter\ is a result of the phase transition which happened after
inflation that has lasted for a certain $e$-folds number. We have derived
two parameters $H_{I}$ and $H_{T}$ that take differents values which could
explain the discrepancy of the Hubble rate parameter measurements. From the
Hybrid potential $V(\phi ,\sigma ),$\ the couplings constants and the bare
masses of the fields must be constrained to satisfy the model we have
proposed in the present paper. Although the physics of reheating is studied
independently on dark energy effects, in this work we consider that dark
energy is a result of\ inflationary\ quantum fluctuations, this provides the
possibility to give new constraints on the reheating parameters $N_{re}\ $%
and $T_{re}$ according to recent data.

\end{document}